\documentclass[graphicx,twocolumn,longbibliography,reprint]{revtex4-1}

\usepackage[version=3]{mhchem} % Formula subscripts using \ce{}
\usepackage[T1]{fontenc}
\usepackage{siunitx}
\usepackage{graphicx}

\usepackage[normalem]{ulem}
\useunder{\uline}{\ul}{}
\begin{document}
\author{Robin J. Dolleman}
\email{R.J.Dolleman@tudelft.nl}
\author{Samer Houri}
\author{Dejan Davidovikj}
\author{Santiago J. Cartamil-Bueno}
\author{Yaroslav M. Blanter}
\author{Herre S. J. van der Zant}
\author{Peter G. Steeneken}
\affiliation{Kavli Institute of Nanoscience, Delft University of Technology, Lorentzweg 1, 2628 CJ, Delft, The Netherlands}
\title{Optomechanics for thermal characterization of suspended graphene}

\begin{abstract}
Thermal properties of suspended single-layer graphene membranes are investigated by characterization of their mechanical motion in response to a high-frequency modulated laser.
A characteristic delay time $\tau$ between the optical intensity and mechanical motion is observed, which is attributed to the time required to raise the temperature of the membrane.
We find, however, that the measured time constants are significantly larger than the predicted ones based on values of the specific heat and thermal conductivity.
In order to explain the discrepancy between measured and modeled tau, a model is proposed that takes a thermal boundary resistance at the edge of the graphene drum into account.
The measurements provide a noninvasive way to characterize thermal properties of suspended atomically thin membranes, providing information that can be hard to obtain by other means.
\end{abstract}

\maketitle
Graphene is a 2-dimensional material with a honeycomb lattice consisting of carbon atoms \cite{geim2007rise}. Amongst  its many unusual properties, its thermal conductance has attracted major attention \cite{pop2012thermal,xu2016phonon}. Extremely high thermal conductivities have been demonstrated up to 5000 W/(m$\cdot$K), well exceeding the thermal conductivity of graphite \cite{balandin2008superior,ghosh2008extremely}. These measurements were performed by Raman spectroscopy, that uses the temperature dependence of the phonon frequency \cite{calizo2007temperature}. By measuring the thermal resistance $R$, which is the local temperature increase $\Delta T$ per unit of heat flux $\Delta Q$, one can employ analytical models of the heat transport to extract the thermal conductivity of graphene $k$. This method allowed demonstration that the thermal conductivity decreases when the number of graphene layers is increased from 2 to 4 \cite{ghosh2010dimensional}. The method has been subsequently improved, for example by better calibration of absorbed laser power \cite{cai2010thermal} or removing parallel conduction paths through the air \cite{chen2010raman}. Also the amplitude ratio between Stokes and anti-Stokes signals has been exploited \cite{faugeras2010thermal} as an alternative to the shift in phonon frequency. As an alternative to Raman measurements, electrical heaters \cite{xu2014length}, pump probe methods \cite{schwarz2016,cabrera2015thermal}, scanning thermal microscopy \cite{yoon2014measuring} 
and temperature sensors \cite{seol2010two} have been used to study heat transport in graphene, demonstrating length dependence of the thermal conductivity \cite{xu2014length} and a reduced thermal conductivity when graphene is supported on silicon dioxide rather than freely suspended \cite{seol2010two}. Different groups have demonstrated a large variety in thermal conductivity of graphene between 600 to 5000 W/(m$\cdot$K) experimentally \cite{balandin2008superior,ghosh2008extremely,cai2010thermal,chen2010raman,faugeras2010thermal,xu2014length,lee2011thermal,dorgan2013high,chen2012thermal,li2014thermal,nika2016thermal} and between 100 to 8000 W/(m$\cdot$K) theoretically \cite{nika2016thermal}
%\cite{nika2009phonon,lindsay2010flexural,lindsay2014phonon,alofi2012phonon,alofi2013thermal,klemens2001theory,nika2009lattice,nika2012anomalous,che2000thermal,osman2001temperature,berber2000unusually,fugallo2014thermal,chen2012thermal,pereira2013divergence,fthenakis2014effect,zhang2011thermal,wei2014mode,lindsay2010diameter,munoz2010ballistic,zhang2012thermal,park2013length,cao2012molecular,serov2013effect,ong2011effect,nika2016thermal},
making the thermal conductance of graphene a debated subject. 

Besides these steady-state studies of the thermal properties of graphene, it is of interest to study its time-dependent thermal properties. This requires measurement of small temperature fluctuations in suspended graphene at frequencies in the MHz range. However, since suspended integration of temperature sensors poses problems and optical techniques for temperature measurement in suspended graphene like Raman do not offer the temperature resolution and frequency bandwidth, direct high-frequency temperature measurement in suspended graphene is difficult.

In this work, it is therefore proposed to use the thermomechanical response of suspended graphene to characterize its thermal properties at MHz frequencies. It is found that the mechanical motion is delayed by a characteristic thermal time constant $\tau$ with respect to the intensity-modulation of the laser that opto-thermally actuates the membrane. This is attributed to the time necessary for heat to diffuse through the system. The optomechanics thus provides a tool for studying the dynamic thermal properties of 2D materials. Interestingly, it is found that the measured values of $\tau$ are much higher than those expected based on literature values for the thermal conductivity $k$, specific heat $c_p$ and density $\rho$ of graphene. Models and measurements of drums of different diameters and on different substrates are analyzed in order to account for the large value of $\tau$. It is found that the long characteristic time is best explained by a large thermal boundary resistance at the edge of the drum.

\begin{figure}
\includegraphics{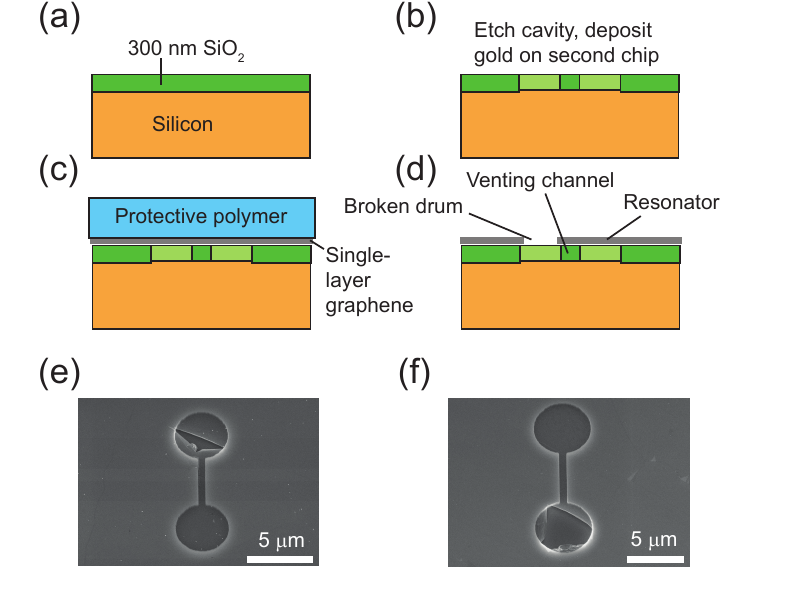}
\caption{Sample fabrication ({a})  Fabrication starts with a silicon die with 280 nm thermally grown silicon dioxide. ({b}) Dumbbell-shaped cavities were etched in the oxide layer. ({c}) Single layer graphene grown by chemical vapor deposition is transferred on both chips with a protective polymer. ({d}) The polymer is dissolved and the sample is dried using critical point drying. This breaks the graphene on one half of the dumbbell, creating a resonator with a venting channel on the other half.  ({e})  Image from a scanning electron microscope (SEM) showing a successful device (3 $\mu$m diameter) with the top part broken and the bottom part whole. ({f}) Successful device (5 $\mu$m diameter) with the bottom part broken and the top part whole. \label{fig:sample}}
\end{figure}
Single-layer graphene resonators are fabricated on top of 300 nm deep dumbbell-shaped cavities (see Fig. \ref{fig:sample}). Two substrates are used, one with the cavities etched in a layer of silicon dioxide and the graphene directly transferred on top. The second substrate is coated with a layer of 5 nm chromium and 40 nm gold before graphene is transferred. This is done to help determine whether the thermal properties of the substrate influence the measured characteristic thermal time. Single layer graphene grown by chemical vapor deposition is transferred over both chips covered with a protective polymer. This polymer is dissolved and the sample is dried using critical point drying (CPD) with liquid carbon dioxide. The fluid forces in this process break one half of the dumbbell, creating a resonator on the other half with a venting channel that lets gas below the membrane escape when the vacuum chamber containing the sample is purged. The graphene is further characterized by Raman spectrocopy and atomic force microscopy (AFM) to confirm it is single layer and contamination levels are low (see Supporting Information S1).

\begin{figure}[b]
\includegraphics{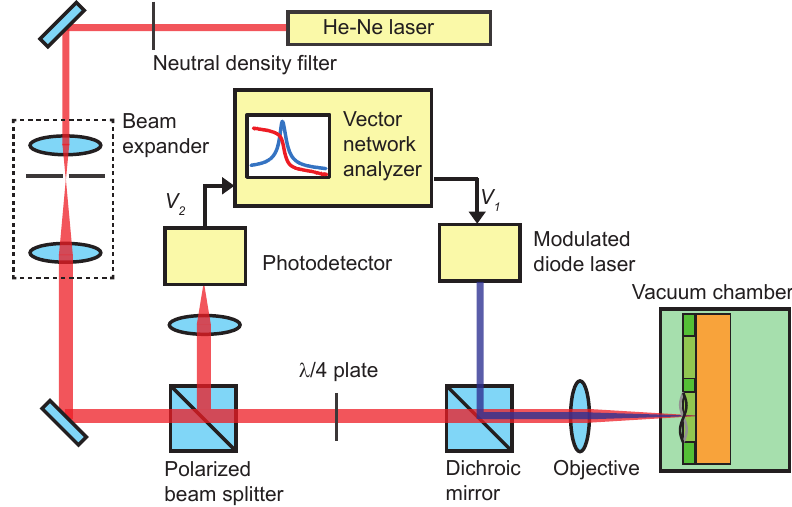}
\caption{Interferometry setup used to actuate and detect the motion of the resonators.\label{fig:setup}}
\end{figure}
The interferometric setup shown in Fig. \ref{fig:setup} is used to actuate the membrane and detect the motion \cite{bunch2007electromechanical,dolleman2015graphene,dolleman2016graphene,castellanos2013single,zande2010large}. In this setup the samples are mounted in a vacuum chamber with optical access. Graphene's motion is detected by cavity optomechanics using a red He-Ne laser, where the suspended membrane acts as moving mirror and the bottom of the cavity as a fixed back-mirror in a low-finesse Fabry-Perot cavity. The intensity of the blue laser is modulated and heats up the membrane, which will deflect due to thermal expansion.  A vector network analyzer (VNA) measures the transmission between the modulation and the signal on the photodetector in a homodyne detection scheme. Frequencies between 100 kHz and 100 MHz can be measured in this setup. All measurements are performed at pressures lower than 0.02 $\mu$bar, reducing gas damping and heat transport through the gas. The red laser power was 1.2 mW incident on the sample and the blue laser power was at 0.36 mW with a large power modulation of 67\% in all experiments.

\begin{figure}[t]
\includegraphics{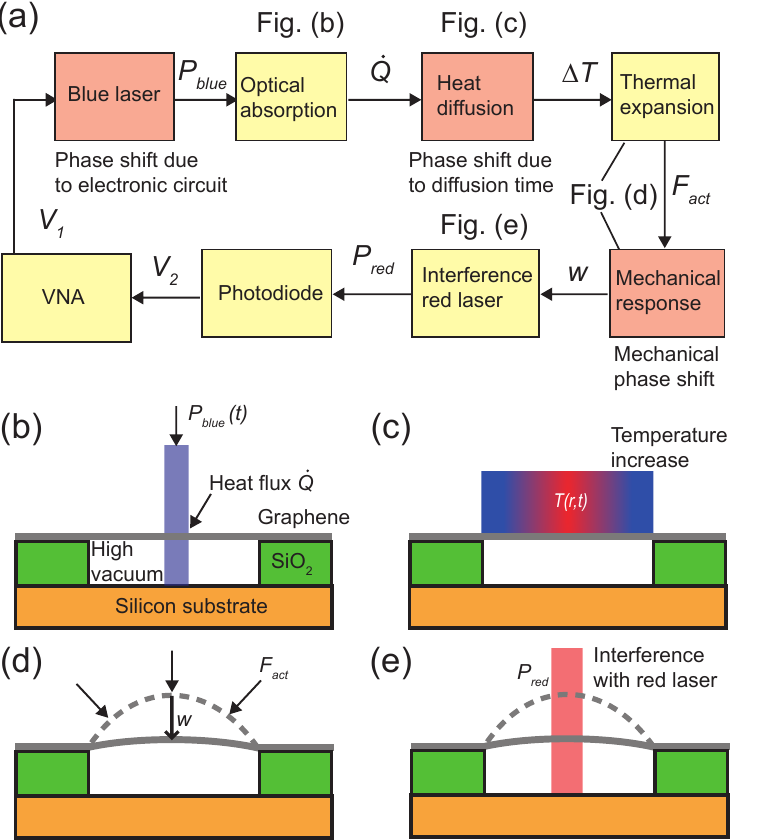}
\caption{Measurement method to determine the characteristic thermal time of suspended graphene resonators. ({a}) Block diagram showing how the deflection signal is transduced using opto-thermal actuation. ({b}) The optical power of the blue laser causes a heat flux in the graphene membrane. ({c}) This heat flux causes a temperature profile in the sample that depends on position and time. ({d}) The increased temperature in the drum causes the membrane to shrink, since graphene has a negative expansion coefficient. This results in the deflection of the graphene. ({e}) The deflection is detected by interference with the red laser, in which the silicon substrate acts as the fixed mirror and graphene as the moving mirror. \label{fig:calibration}}
\end{figure}

Here we identify the potential source for time delay between the modulation of the blue laser and the mechanical response in the measurement setup. The block diagram in Fig. \ref{fig:calibration}({a}) identifies the elements and processes that play a role in actuation and detection of the membrane's motion. The modulated intensity of the blue laser is absorbed in the graphene, generating a virtually instantaneous heating power (Fig. \ref{fig:calibration}({b})) since photoexcited carriers in graphene lose their energy to phonons on timescales of a few picoseconds \cite{wang2010ultrafast}. The generated heat will increase the temperature of the membrane and flow toward the substrate, resulting in a time-dependent temperature increase of the membrane, where the temperature is delayed with respect to the heating power (Fig. \ref{fig:calibration}({c})). The temperature increase causes thermal expansion forces that deflect the membrane (Fig. \ref{fig:calibration}({d})). At frequencies far below the resonance frequency the motion will be in-phase with the thermal expansion force, especially since the quality factor of the resonanator is typically higher than 100. The intensity modulation of the red laser due to interference effect that is used to detect the motion (Fig. \ref{fig:calibration}({e})) can be regarded as instantaneous and will not cause a delay. The measurements are corrected for other delays, related to delays in the instruments (VNA, photodiode) and light path delays, using a calibration procedure discussed in the Supporting Information S3.

It is thus concluded that in the frequency range below the mechanical resonance, the delay between optical actuation and deflection in Fig. \ref{fig:calibration} is nearly completely due to the delay between heating power and temperature. A thermal system with a single time constant $\tau$, driven by an ac heating power $P_{ac}e^{i \omega t}$ can be described by the heat equation:
\begin{equation}
  \frac{\mathrm{d} \Delta T}{\mathrm{d} t}+\frac{1}{\tau} \Delta T=\frac{P_{ac}}{C} e^{i\omega t},
  \end{equation}
   where $\Delta T$ is the temperature difference with respect to the steady-state temperature, $C$ is the thermal capacitance and $\tau=RC$ is the thermal $RC$ product. At frequencies significantly below the mechanical resonance frequency, the thermal expansion induced amplitude $z=\alpha \Delta T$ is proportional to temperature by an effective thermal expansion coefficient $\alpha$. Solution of the heat equation gives:
\begin{equation} 
z_\omega e^{i \omega t}=\alpha R P_{ac} \frac{ e^{i\omega t}}{i \omega \tau + 1}.
\label{heateq}
\end{equation}
In section S6 of the Supporting Information a full derivation of the complex amplitude $z_{\omega}$, including the mechanical damping and inertia effects, is given based on derivations by Metzer et al' \cite{metzger2008optical,metzger2004cavity,favero2007optical}. This equation will be used to fit the experimental data, with the parameters $B=\alpha R P_{ac}$ and $\tau$.

\begin{figure}[t]
\includegraphics{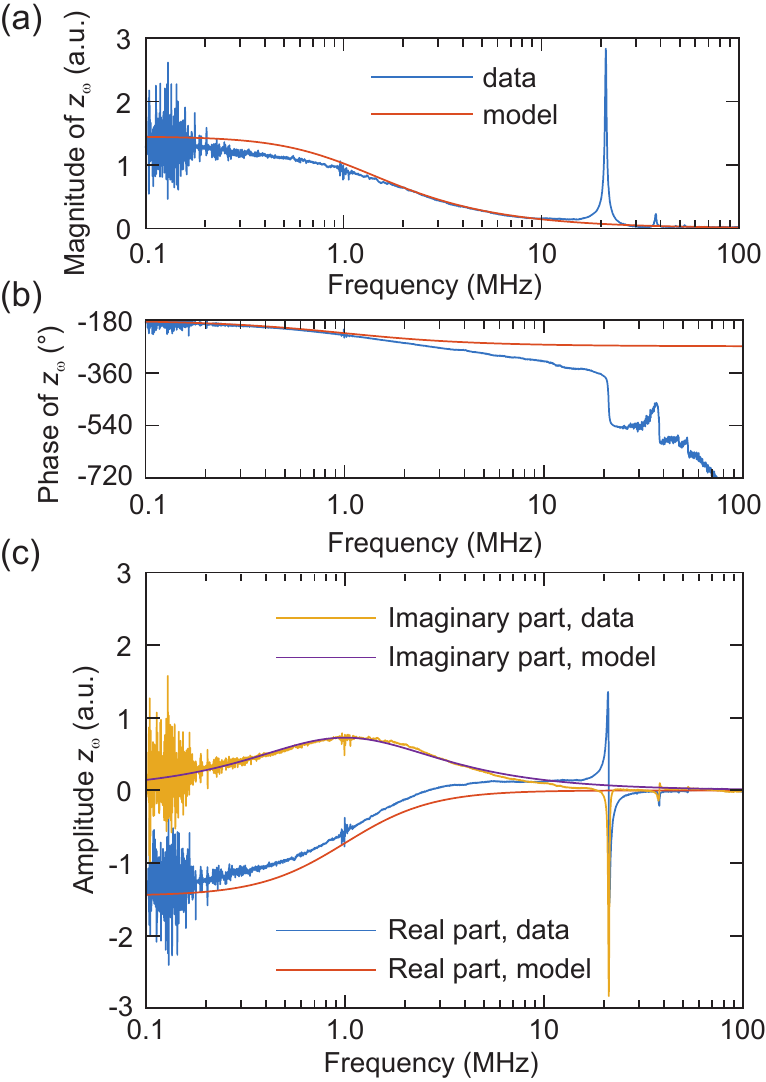}
\caption{Typical measured frequency response function ({a}) Magnitude and ({b}) phase of the VNA signal after calibration, showing a decrease in magnitude and a phase shift well before the resonance frequency at 22 MHz. ({c}) Real and imaginary parts of the signal, with a fit from eq. \ref{heateq} to the imaginary part. The expected real part from this model is also shown in this plot. \label{fig:phase}}
\end{figure}
An example of the measured magnitude and phase of the deflection for a resonator with a diameter of 5 $\mu$m on a cavity in silicon dioxide is shown in Fig. \ref{fig:phase}({a}) and \ref{fig:phase}({b}) respectively. In the 0.1 to 10 MHz range, the response is frequency dependent with a decrease in magnitude as the frequency increases. Also a phase delay is observed that increases as function of frequency.  Note, that the measured phase at low frequencies is not 0, but 180 degrees. This is attributed to the small offset in the deflection that the graphene membrane has, in some membranes this was reversed in sign (indicated by 0 degrees phase at low frequencies) as shown in the Supporting Information S2. Figure \ref{fig:phase}({c}) shows a measurement result which is split into a real and an imaginary part. The imaginary part of the amplitude $z_{\omega}$ can be fit by eq. \ref{heateq}, resulting in a value of characteristic delay time of $\tau = 159 $ ns, with a clearly observable maximum at radial frequency $\omega=1/\tau$. The real part of eq. \ref{heateq}, with the same $B$ and $\tau$, is shown in Fig. \ref{fig:phase}(c) showing a small offset with respect to the data. The offset is attributed to optical cross-talk from the modulated blue laser, which can reach the photodetector despite the optical isolation.  The same effect causes the difference between the model and the magnitude and phase of the amplitude response (see Figs. \ref{fig:phase}(a) and \ref{fig:phase}(b)).

\begin{figure*}[t]
\includegraphics{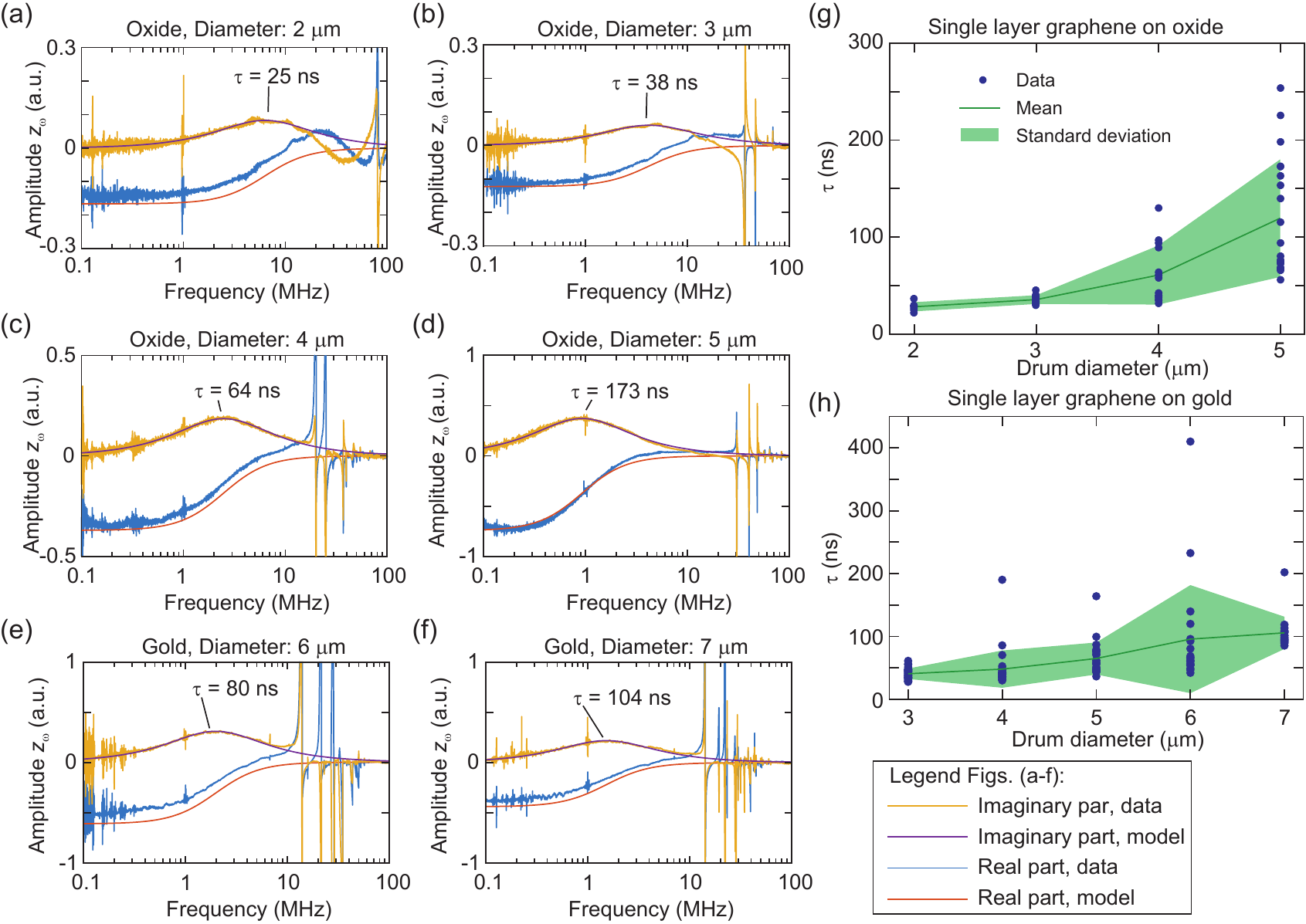}
\caption{Measured characteristic times compared for different diameters. ({a}-{f}) Typical measurements for different diameters and substrates. Optical cross-talk is visible as an offset in the amplitude in the real part of approximately 0.05 to 0.1, which is less visible for samples that show larger amplitudes. In general eq. \ref{heateq} describes the behavior well between 0.1 and 10 MHz.  ({g}) $\tau$ as function of diameter for single layer graphene suspended on a silicon dioxide substrate, showing that both $\tau$ increases as well as the spread in $\tau$. ({h}) $\tau$ as function of diameter for single layer graphene drums suspended on a gold substrate. \label{fig:diameter}}
\end{figure*}
Figs. \ref{fig:diameter}({a}-{f}) show typical measurement results for drums with diameters from 2-7 $\mu$m. The optical cross-talk is significantly higher in the substrates coated with gold, which is visible as a larger offset in the real part. The resulting values of $\tau$ as function of diameter for both the silicon dioxide substrate and the gold substrate are plotted in Figs. \ref{fig:diameter}({g},{h}), respectively. A trend is observed where $\tau$ increases as function of diameter. No significant correlation between $\tau$ and fundamental resonance frequency $\omega_0$ was found as shown in the Supporting Information S4, suggesting weak dependence of $\tau$ on strain.

The measured time constants in this work are significantly larger than expected based on the intrinsic properties of graphene. For example, Barton et al. \cite{barton2012photothermal} use an expression that estimates the time constant based on the thermal properties of graphene:
\begin{equation}\label{eq:inttau}
\tau = \frac{a^2 \rho c_p}{2 k},
\end{equation}
where $a$ is the membrane radius, $\rho$ the density of graphene, $c_p$ specific heat and $k$ the thermal conductivity. Using approximate values $c_p = 600$ J/(kg$\cdot$K) (calculated in the Supporting Information S7), $k = 2500$ W/(m$\cdot$K) and $\rho = 2300$ kg/m$^3$ we obtain $\tau = 0.3$ ns for a 2 micron drum and $\tau = 2$ ns for a 5 micron drum. The observed values of $\tau$ range between 25 to 250 ns, which is one to two orders of magnitude larger than those predicted by eq. \ref{eq:inttau}. Even if the most extreme values for $c_p$ and $k$ are used, eq. \ref{eq:inttau} gives a lower $\tau$ than measured. The theoretical limit for $c_p$ is given by the Petit-Dulong law ($c_p = 2100$ J/kg/K), and the lowest experimental literature value for $k$ is 600 W/(m$\cdot$K) \cite{faugeras2010thermal}. It thus appears that eq. \ref{eq:inttau} cannot account for the experimental $\tau$. 

Therefore we consider the possibility that the thermal conduction is limited by the substrate that supports the graphene resonator. In order to investigate this, we compare the results obtained on gold-coated and uncoated substrates. It is found that the $\tau$ on the different substrates are similar (Fig. \ref{fig:diameter}({g}-{h})), despite the much higher thermal conductivity of the gold-coated substrate. It is thus concluded that substrate effects are not responsible for the observed value of $\tau$. This conclusion is consistent with finite element simulations (Supporting information S5) of the system. 

It is well known that a thermal resistance can be present at the interface between two solids \cite{pollack1969kapitza,stevens2007effects,swartz1989thermal,peterson1973kapitza,costescu2003thermal,lyeo2006thermal}. This effect is called interfacial thermal (or Kapitza) resistance and is caused by differences in the phonon velocities, which leads to scattering that limits the phonon transport across the interface. Several works have predicted interfacial resistances in graphene using molecular dynamics simulations \cite{pei2012carbon,xu2014interfacial}. Between suspended and supported graphene a value of the boundary conductance of $2\times 10^{10}$ W/(K$\cdot$m$^2$) was reported \cite{xu2014interfacial}. Also, grain boundaries in graphene have been shown to cause an interfacial thermal resistance \cite{isacsson2016scaling}. Below we argue that an interfacial thermal resistance between supported and suspended graphene could account for the unexpectedly long thermal  delay times we measured.

\begin{figure*}
\includegraphics{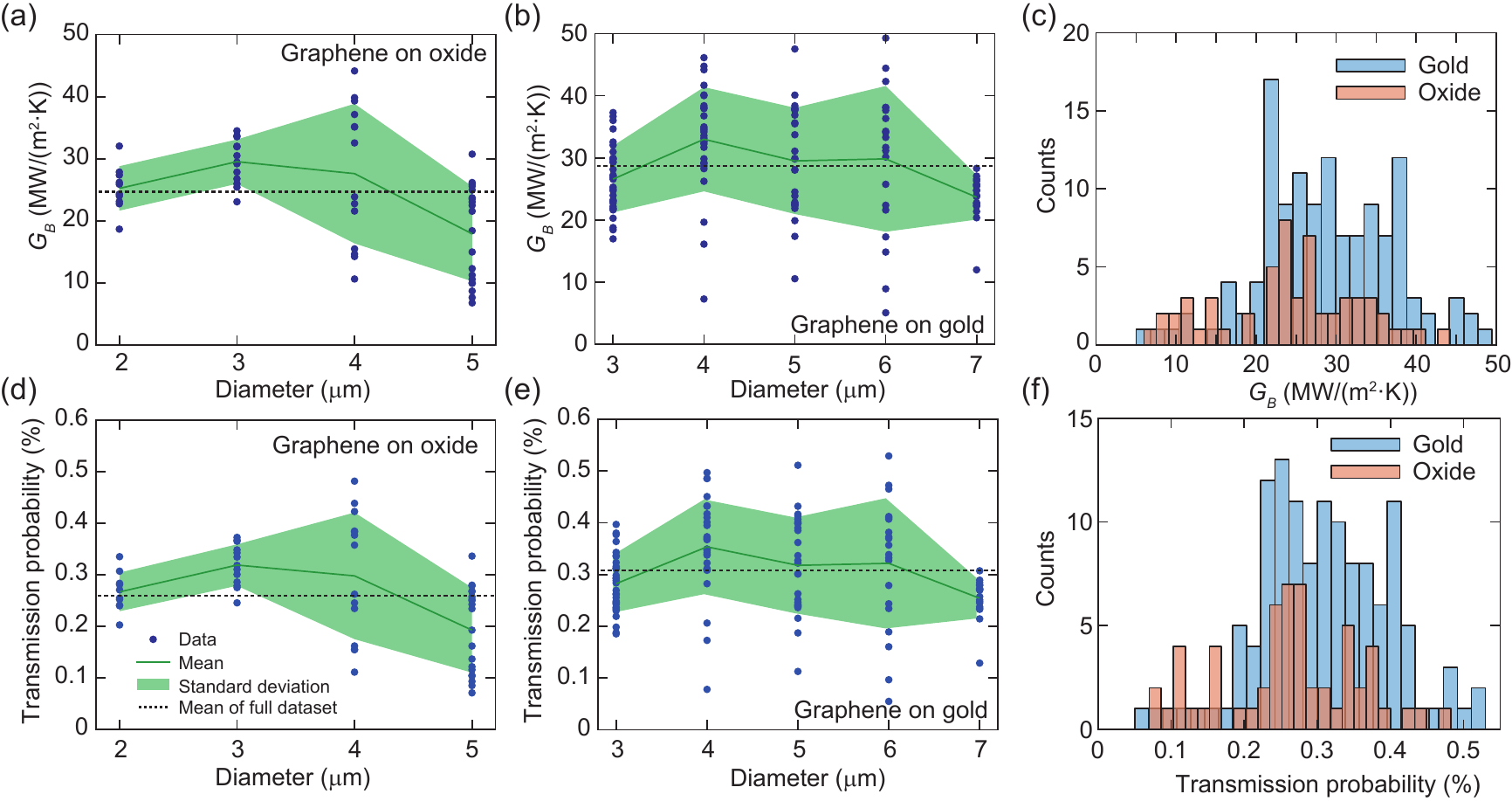}
\caption{Properties of the thermal boundary extracted from the measurements. ({a}) Thermal boundary conductance $G_B$ extracted using eq. \ref{eq:tauexpression} as function of diameter for the oxide sample and ({b}) for the sample covered with gold. ({c}) Histogram comparing all measured conductances showing the similar distributions between both the oxide and gold dataset. ({d}) Transmission probability from the supported to the suspended part of graphene for the oxide and ({e}) for the gold sample. ({f}) Histogram showing all obtained transmission probabilities. \label{fig:TBC}}
\end{figure*}
The boundary resistance will cause the formation of a temperature discontinuity at the interface between suspended and supported graphene that can be modeled by Fourier's law \cite{stevens2007effects}:
\begin{equation}
Q_B = \frac{T_{\mathrm{sus}} - T_{\mathrm{sup}}}{R_B} \equiv G_B (T_{\mathrm{sus}} - T_{\mathrm{sup}}),
\end{equation}
where $Q_B$ is the boundary heat flux, $T_{\mathrm{sus}}$ the temperature in the suspended part of the graphene and $T_{\mathrm{sup}}$ temperature of the supported part. $R_B$ is the thermal boundary resistance and $G_B$ is the thermal boundary conductance. In order to estimate $G_B$ we use a thermal RC model, where the thermal time $\tau$ is given by de product of the heat capacity of suspended graphene $C$ and the thermal resistance $R$. It is assumed that $R$ is dominated by the interfacial thermal resistance $R_B$, such that $\tau$ becomes independent of $k$ of graphene:
\begin{equation}
C = c_p \rho h_g \pi a^2,
\end{equation}
\begin{equation}
R = (G_B h_g 2 \pi a)^{-1},
\end{equation}
where $h_g$ is the thickness of single layer graphene. Combining both expressions yields for the thermal time $\tau$:
\begin{equation}\label{eq:tauexpression}
\tau = \frac{ \rho c_p a}{2 G_B}
\end{equation}

Using eq. \ref{eq:tauexpression} the thermal boundary conductance ($G_B = \rho c_p a / 2 \tau$)  is derived from the measurements of $\tau$ as shown in Fig. \ref{fig:TBC}({a}), \ref{fig:TBC}(b) and \ref{fig:TBC}({c}). This shows that the value of the thermal boundary conductance lies around 30 MW/(m$^2 \cdot$K). For the purpose of extracting $G_B$ the derived value $c_p= 600$ J/(kg$\cdot$K) is used and a density $\rho=2300$ kg/m$^3$. Equation \ref{eq:tauexpression} has been verified using finite element simulations that include a thermal boundary conductance, confirming the validity of neglecting the heat conductance $k$ (Supporting Information S5).

In order to relate the derived value of $G_B$ to the phonon transmission probability across the interface, the following expression is derived in Supporting Information S7:
 \begin{eqnarray}\label{eq:kapitzaR}
  \tau = \frac{ \rho c_p a}{2 G_B} =  \frac{a}{2} \frac{ \dfrac{1}{c_{1l}^2} +  \dfrac{1}{c_{1t}^2} + \dfrac{\pi \hbar^2}{3 \zeta(3) A_{uc} k_B^2 T^2}}{\dfrac{\bar{w}_{1l}}{c_{1l}} + \dfrac{\bar{w}_{1t}}{c_{1t}} +\dfrac{\pi \hbar^2 \bar{w}_{1z} c_{1z}}{\zeta(3) A_{uc} k_B^2 T^2 }}
 \end{eqnarray} 
Here $c_{1j}$ is the velocity of the $j$-th phonon mode, $j=z$ for the flexural (ZA), $j=l$ for longitudinal (LA) and $j=t$ for the transverse (TA) mode. The number 1 corresponds to the suspended material. $\bar{w}_{1j}$ is the integrated transmission probability (the sum over each possible angle of incidence) of phonons over the interface, $k_B$ is Boltzmann constant, $T$ is temperature, $\hbar$ the reduced Planck's constant and $A_{uc}$ is the area of the unit cell of graphene. 

By using eq. (\ref{eq:kapitzaR}), an average phonon transmission probability $\bar{w}$ is plotted in Figs. \ref{fig:TBC}(d), \ref{fig:TBC}(e) and \ref{fig:TBC}(f) corresponding to the boundary conductances in Figs. \ref{fig:TBC}(a), \ref{fig:TBC}(b) and \ref{fig:TBC}(c). The average phonon transmission probability is found to be $\bar{w} = 0.3 \pm 0.2$ \%. Potential mechanisms that limit $\bar{w}$ include phonon interface scattering due to differences in phonon propagation velocities, boundary roughness \cite{wen2009specular} and kinks \cite{evans2013reflection} due to graphene edge adhesion \cite{bunch2008impermeable}. Further experimental and theoretical study of heat transport across the edge between suspended and supported graphene is needed to clarify the microscopic origins of these observations.

To summarize, a dynamic optomechanical method to measure transient heat transport in suspended graphene is demonstrated. The method does not require electrical contacts, which allows high-throughput characterization of arrays of devices. The method is used to characteristic the thermal time $\tau$ of many graphene membranes. It is found that $\tau$ is a function of diameter and its value is much larger than expected based on existing models. Measurements on gold-coated and uncoated silicon dioxide samples show similar results, showing that $\tau$ cannot be attributed to the substrate. A potential cause for the large values of $\tau$ is the presence of an interfacial thermal resistance between the suspended and supported graphene. From the measurements we determine that a thermal boundary conductance with values of $30 \pm 20$  MW/(m$^2 \cdot$K) can account for the measurements, corresponding to a low phonon transmission probability on the order of 0.3\%. 

\section*{Acknowledgments}
We thank Applied Nanolayers B.V. for supply and transfer or single-layer CVD graphene on our substrates. We also acknowledge useful discussions with Teun Klapwijk, Debadi Chakraborty, Daniel Ladiges and John Sader.
Further, we thank the Dutch Technology Foundation (STW), which is part of the Netherlands Organisation for Scientific Research (NWO), and which is partly funded by the Ministry of Economic Affairs, for financially supporting this work.
The research leading to these results also received funding from the European Union's Horizon 2020 research and innovation programme under grant agreement No 649953 Graphene Flagship and this work was supported by the Netherlands Organisation for Scientific Research (NWO/OCW), as part of the Frontiers of Nanoscience program.

\onecolumngrid
\pagebreak
~
\\
\pagebreak 

\section*{Supplemental Information}
\setcounter{equation}{0}
\setcounter{figure}{0}
\setcounter{table}{0}
\makeatletter
\renewcommand{\theequation}{S\arabic{equation}}
\renewcommand{\thefigure}{S\arabic{figure}}
\renewcommand{\thepage}{S\arabic{page}}  

\title{Supplemental Information: Optomechanics for thermal characterization of suspended graphene}
\author{Robin J. Dolleman}
\email{R.J.Dolleman@tudelft.nl}
\author{Samer Houri}
\author{Dejan Davidovikj}
\author{Santiago J. Cartamil-Bueno}
\author{Yaroslav M. Blanter}
\author{Herre S. J. van der Zant}
\author{Peter G. Steeneken}
\affiliation{Kavli Institute of Nanoscience, Delft University of Technology, Lorentzweg 1, 2628CJ, Delft, The Netherlands}

\maketitle

\section{Additional Experimental results and methods}

\subsection*{S1: Graphene characterization}
\begin{figure}[h]
\includegraphics{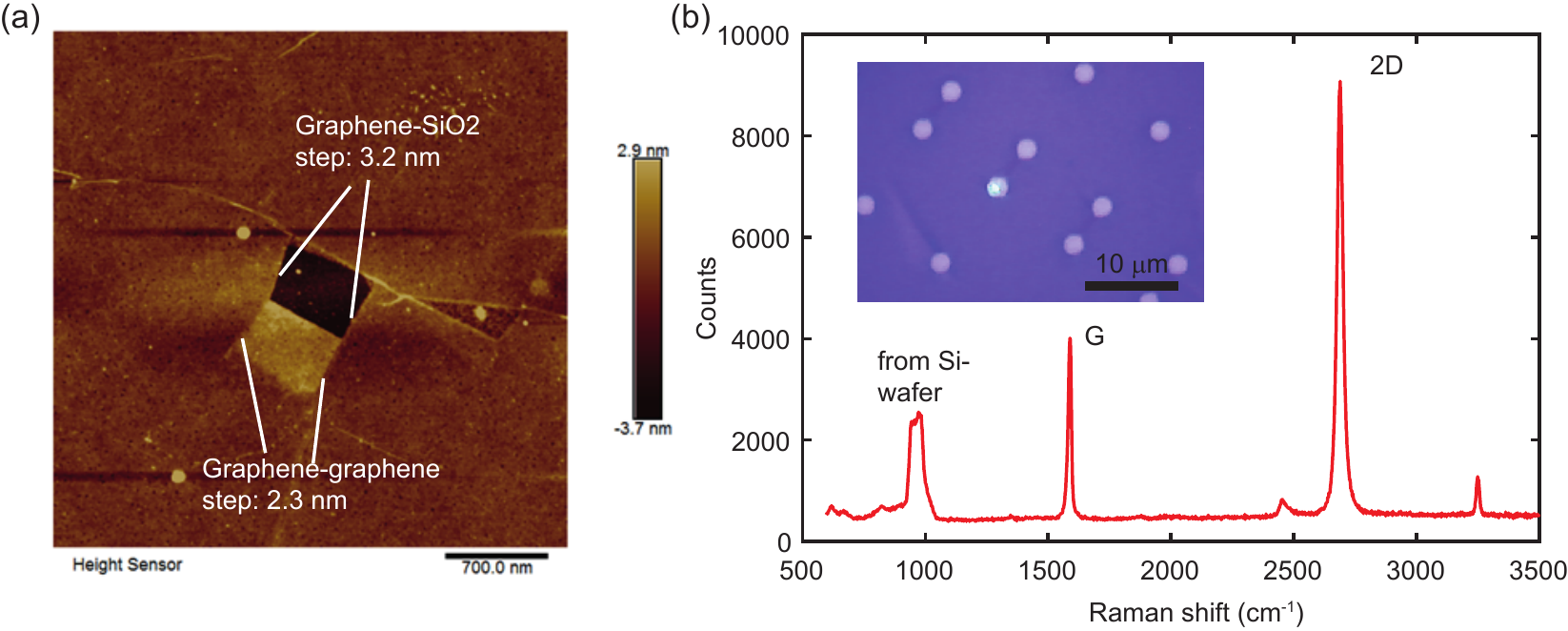}
\caption{({a}) Tapping-mode atomic force microscopy measurement of the thickness of the graphene. Step heights were measured on the marked locations. ({b}) Raman spectroscopy measurements. The 2D:G peak height ratio of 2:1 is a clear indication that the graphene is single-layer. Also the absense of a D peak suggests that the graphene has low defect density. The measurement was performed on a 2 micron diameter suspended drum (inset).\label{fig:characterization}}
\end{figure}
The graphene was examined for contamination and defects using atomic force microscopy (AFM) and Raman spectroscopy. Figure \ref{fig:characterization}(a) shows a height profile obtained using tapping mode AFM on graphene supported on silicon dioxide. A small part of the graphene sheet is broken and folded over, allowing to measure the step heights between graphene on graphene and graphene on silicon dioxide. The tapping mode AFM can overestimate the step height when measuring 2D materials \cite{nemes2008anomalies}. This inaccuracy is illustrated by the disparity between the graphene-oxide and the graphene-graphene step height. However the folded-over part of the graphene allows us to say that no more than $\sim$ 1.9 nm of uniform polymer contamination could be present on the membrane. This value represents a worst-case scenario and it is very likely that the contamination is much lower. The ratio between the 2D and G peak of 2:1 in the Raman spectroscopy (see Fig. \ref{fig:characterization}(b)) and the absence of a D-peak is consistent with high-quality single layer graphene with low contamination levels \cite{ferrari2006raman}. 

\subsection*{S2: Example of measurement with reversed phase}
\begin{figure}[h!]
\includegraphics{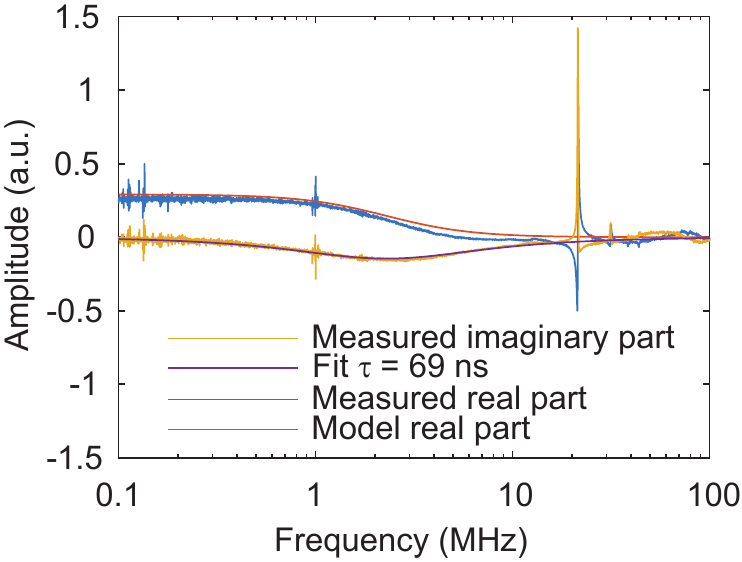}
\caption{Real and imaginary part of the frequency response of the measured drum. Note, that the sign of the amplitude changed with respect to the examples in the main text. \label{fig:spectral}}
\end{figure}
In the main text it is mentioned that the response of the drum can show either a 180 degrees phase shift between laser power and mechanical motion at low frequencies, but also a 0 degrees phase is possible. This is equivalent to a sign change of the real and imaginary part of the amplitude. Although 180 degrees is usually observed in our measurements, occasionally a 0 phase is observed as shown in Fig. \ref{fig:spectral}. Here, the sign of real and imaginary part have changed with respect to the examples in the main text. The cause of this difference in phase are small asymmetries in the initial state which will cause either upward or downward motion due to the thermal expansion.

\subsection*{S3: Calibration procedure}
\begin{figure}[h!]
\includegraphics{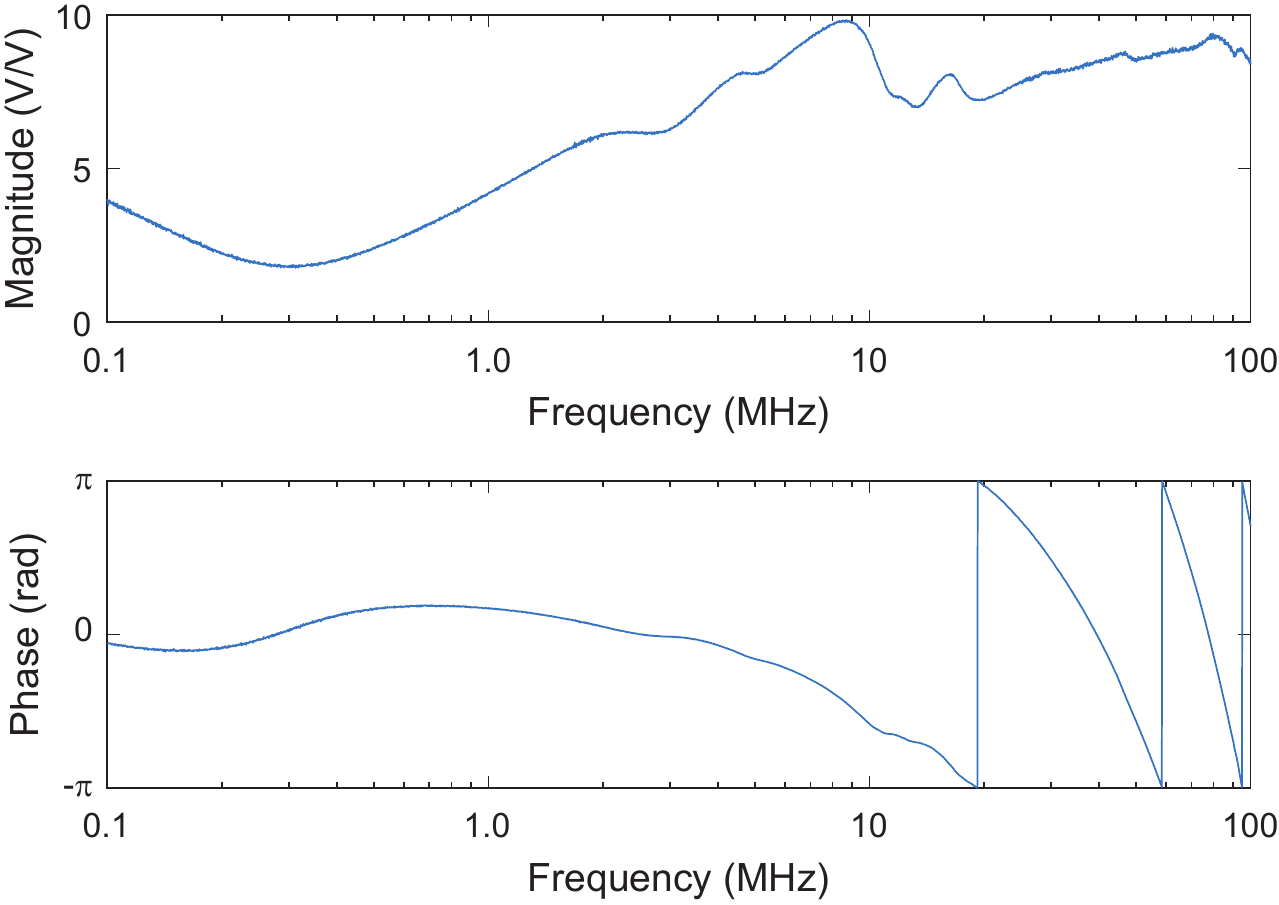}
\caption{Magnitude and phase of the photo-diode signal obtained when the blue laser is directly aimed at the photo-diode. \label{fig:calblue}}
\end{figure}
In order to correct the intrinsic phase shifts in our measurement setup, we directly point the blue laser to the photodiode to obtain an calibration curve for our system (Figure \ref{fig:calblue}). This can be corrected by deconvolution of the measured response with this calibration curve, which is done by expressing the blue laser modulation parameters as a frequency-dependent phasor $ \varepsilon_{\omega}$. Since the calibration was taken at discrete frequencies, a cubic interpolation was used to make sure the frequencies match the ones from the measurement that needs to be corrected. Now one can deconvolve the measured frequency response function $f_{\omega}$ using:
\begin{equation}
|f_{\omega,\mathrm{corr}}| = \frac{|f_{\omega}|}{|\varepsilon_{\omega}|},
\end{equation}
\begin{equation}
\angle f_{\omega,\mathrm{corr}} = \angle f_{\omega} - \angle \varepsilon_{\omega},
\end{equation}
where $f_{\omega,\mathrm{corr}}$ is the corrected frequency response function of our measurement.

\subsection*{S4: Correlation between resonance frequency and $\tau$}
\begin{figure}[h!]
\includegraphics{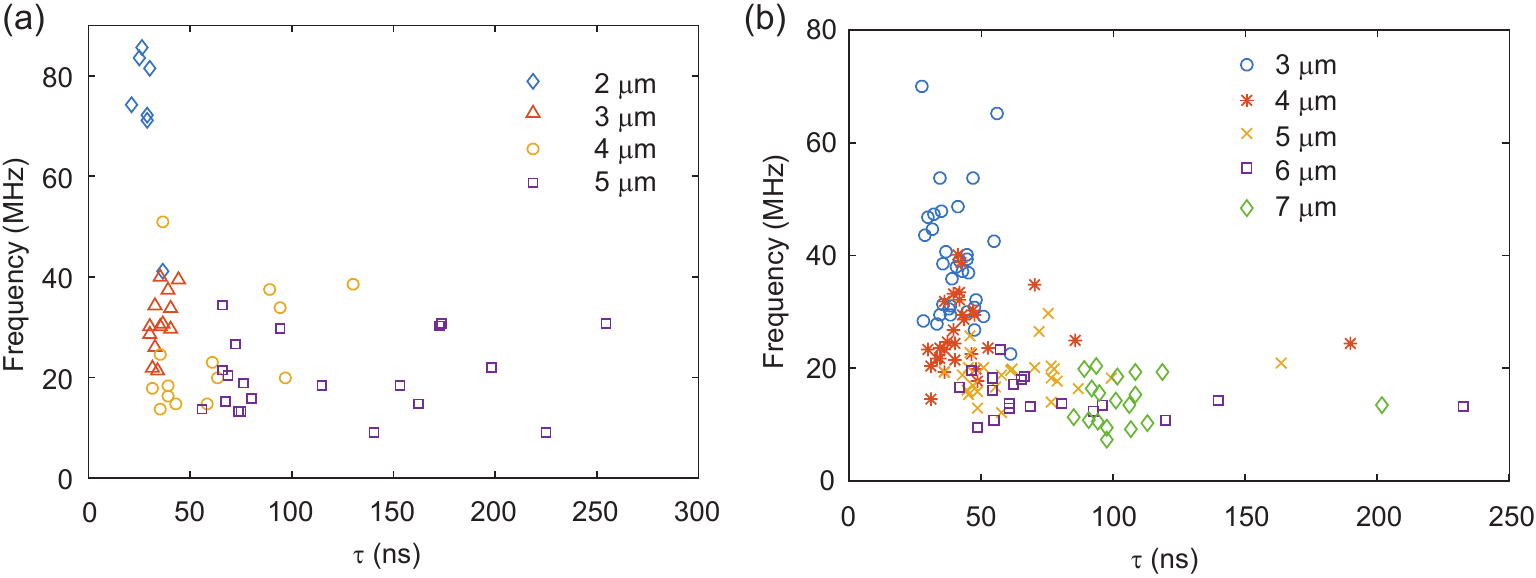}
\caption{(a) Scatter plot between resonance frequency and $\tau$ for the uncoated. (b) Gold-coated sample (one data point with $\tau = 410$ ns, 6 micron diameter and $f=13.6$ MHz not shown here). \label{fig:scatterplot}}
\end{figure}
Figure \ref{fig:scatterplot} shows scatter plots between the resonance frequency and characteristic thermal time $\tau$ extracted from the measurements. Since both the resonance frequency and characteristic thermal time are correlated to diameter, correlations between the two variables should only be determined for the same diameter. We found low correlations close to zero with outliers at -0.24 for the 6 micron diameter drums on gold and 0.14 for the 5 micron diameter drums on gold. The low correlations and the low agreement between different diameters suggests that transient thermal transport is not strongly related to the strain present in the graphene resonators.

\section{Additional modeling results}

\subsection*{S5: Finite element simulations of graphene on a silicon dioxide substrate}
\begin{figure}
\includegraphics{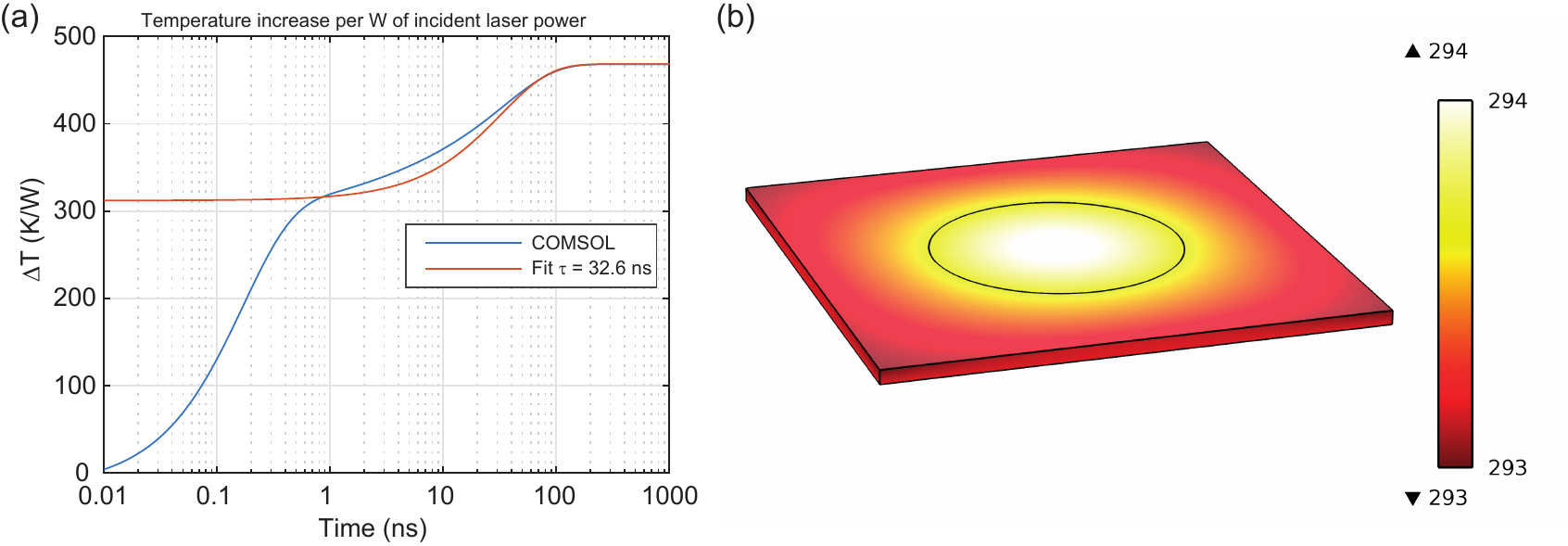}
\caption{({a}) Average temperature of the graphene membrane as function of time obtained by COMSOL, simulating a graphene membrane ($\rho = 2100$ kg/m$^3$, $C = 700$ J/kg/K, $k = 2500$ W/m/K and thickness 0.335 nm) on top of silicon dioxide ($\rho = 2200$ kg/m$^3$, $C = 730$ J/kg/K, $k = 1.4$ W/m/K and thickness 300 nm) with a thermal conductance $h_{\mathrm{T}} = \num{8.33e7}$ W/m$^2$/K between the graphene and the oxide. ({b}) Simulated oxide layer and drum with diameter in thermal equilibrium and 1 mW of laser power. The simulation suggests significant temperature increase outside the drum. \label{fig:COMSOL1}}
\end{figure}

 In order to examine the impact of the silicon dioxide substrate on the heat transport, we use COMSOL Multiphysics to model the graphene on top of the cavity and estimate the delay function $h(t)$. A simulation result of $h(t)/P_{\mathrm{laser}}$, where $P_{\mathrm{laser}}$ is the incident laser power (assuming $2.3\%$ absorption of optical power), can be seen in Fig. \ref{fig:COMSOL1}. This simulation predicts that the heat transport is more complex than expected. A very fast increase in temperature is observed with a time constant that is in the order of 0.5 ns. This is followed by a much slower exponential increase in temperature, which can be fitted with a single exponential to obtain a time constant of 32.6 ns. The fast time constant should not be observed in our measurement, since the cut-off frequency $2 \pi \omega_{c,\mathrm{fast}} = 1/\tau_{\mathrm{fast}} \approx 320$ MHz is much larger than the bandwidth in our measurements. The slow time constant can be observed in our measurement, since the cut-off is in a measurable frequency range and lower than the resonance frequency.  This could be the thermal relaxation time found in the measurements. 
 
\begin{figure}
\includegraphics{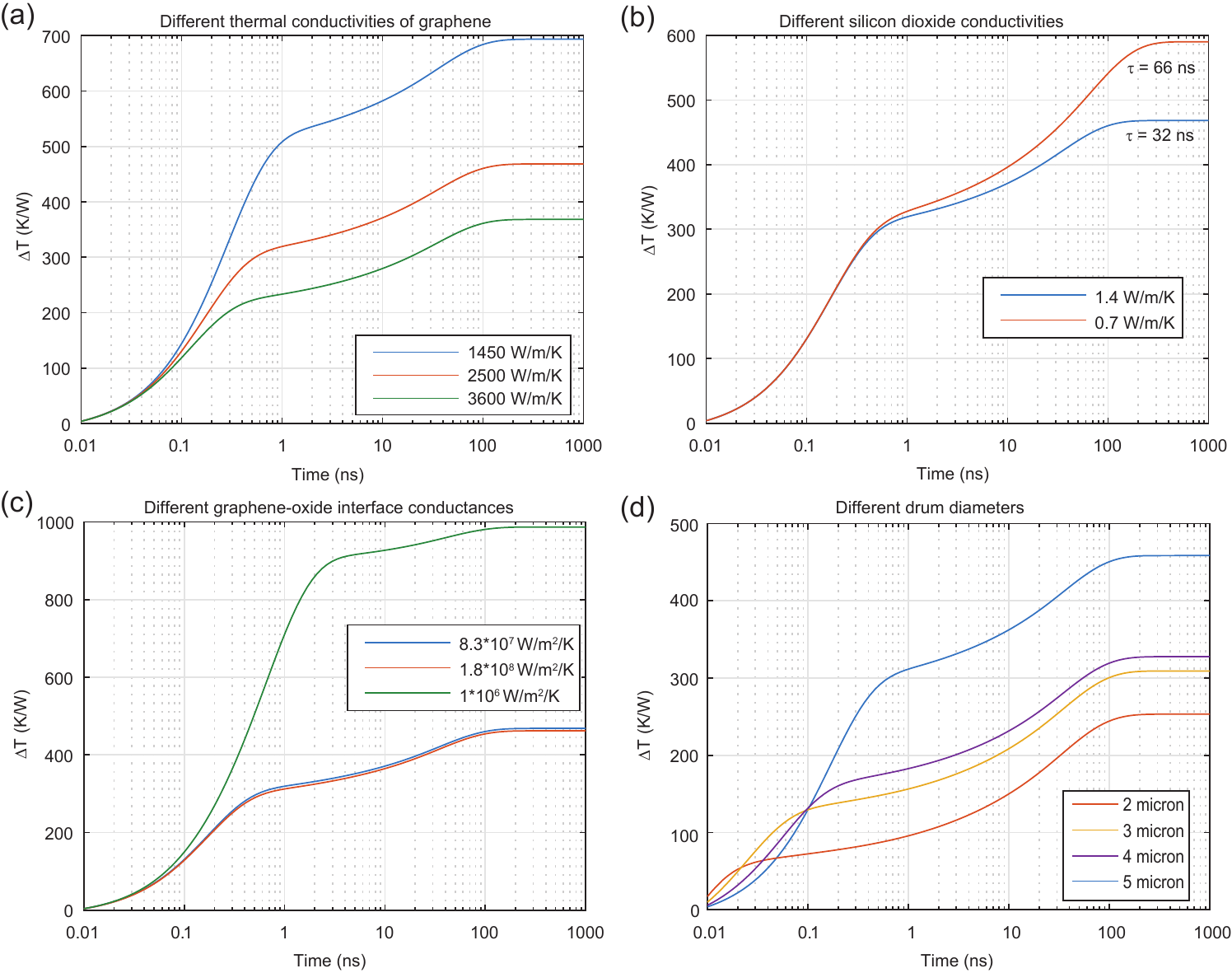}
\caption{({a}) $h(t)$ for different thermal conductivities of graphene. It can be seen that $\tau_{\mathrm{fast}}$ shows some variation, but the slow thermal time constant does not change significantly. The fits of $\tau_{\mathrm{slow}}$ are between 31.9 ns and 32.7 ns. ({b}) $h(t)$ for different thermal conductivities of the silicon dioxide layer. $\tau_\mathrm{slow}$ is highly affected by this value: changing the value from 1.4 W/m/K to 0.7 W/m/K doubles $\tau_{\mathrm{slow}}$ to 66 ns. ({c}) $h(t)$ for different thermal contact resistances. First two values are within ranges typically observed in literature \cite{chen2009thermal}. $\tau_{\mathrm{slow}}$ is again hardly affected, with values between 31.6 ns and 31.9 ns. The last value is an extreme example that is much lower than found in literature. \cite{cai2010thermal,chen2009thermal,mak2010measurement,freitag2009energy} ({d}) $h(t)$ for different drum diameters, here only the fast time constant shows large variations, but the slow time constant is not significantly affected.\label{fig:COMSOLthermalk}}
\end{figure}

Figure \ref{fig:COMSOLthermalk} shows $h(t)$ for different material parameters. From fig. \ref{fig:COMSOLthermalk}(a) we conclude that changes in the thermal conductivity or specific heat of the graphene membrane affect the fast time constant $\tau_{\mathrm{fast}}$, the slow time constant $\tau_{\mathrm{slow}}$ remains unchanged. We conclude that the observed time constant $\tau$ in our measurements does not depend on the properties of graphene itself. For the thermal contact resistance between the graphene and silicon dioxide interface shown in fig. \ref{fig:COMSOLthermalk}(b), we draw the same conclusion.

Figure \ref{fig:COMSOLthermalk}(c) shows a different situation; if the thermal properties of the silicon dioxide layer are changed, the fast time constant remains unchanged. However, the slow time constant changes a lot, from 32 ns to 66 ns if the thermal conductivity is changed from 1.4 W/m/K to 0.7 W/m/K. We therefore conclude that the slow time constant in the simulation depends only on the properties of the substrate. Due to the poor thermal properties of this layer, it takes much longer to reach thermal equilibrium than expected if only the graphene itself is considered. Figure \ref{fig:COMSOLthermalk}(d) shows the diameter dependence on $h(t)$, again the slow time constant is hardly affected, therefore this model does not account for the diameter dependence in our measurement.  Since the slow time constant modeled here only depends on the properties of the substrate, it should significantly change value when the silicon dioxide is replaced with gold. Since the values simulated here are not diameter dependent, this model shows that the experimental observations cannot be explained by the thermal properties of the substrate.

\begin{figure}
\includegraphics{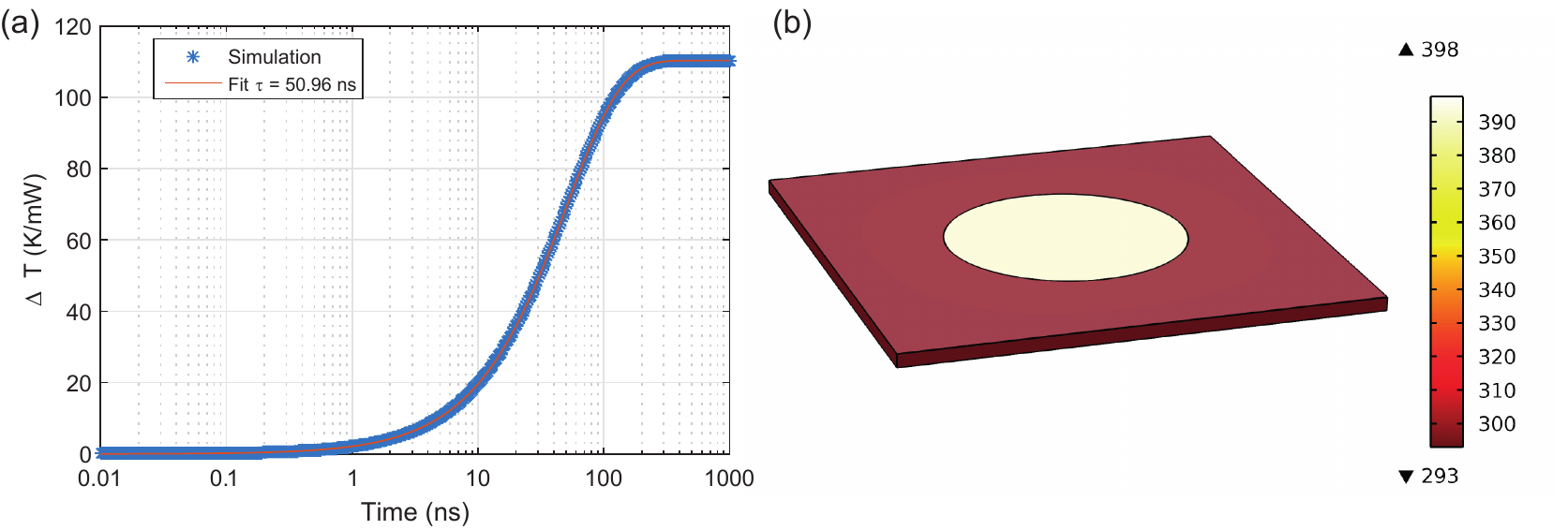}
\caption{Simulation with a boundary conductance $G_B =$ 40 MW/(m$^2 \cdot$ K) between suspended and supported graphene. (a) $h(t)$ showing a single exponential function. (b) Temperature profile of the drums, showing that the temperature is uniform and the materials outside the drum do not change temperature significantly. \label{fig:withboundcond}}
\end{figure}
Figure \ref{fig:withboundcond} shows a simulation with identical parameters as in Fig. \ref{fig:COMSOL1} with the addition of a limited thermal boundary conductance $G_B =$ 40  MW/(m$^2 \cdot$ K). The interfacial thermal resistance dominates the heat transport in this situation as illustrated by the uniform temperature in the drum. The materials outside the drum do not raise in temperature significantly. This validates the simple model in the main part of this work, where only the boundary conductance is considered. 

\subsection*{S6: Derivation of expression for motion of drum actuated by modulated laser}
The derivations below follow the methods used by Metzger et al.\cite{metzger2008optical,metzger2004cavity,favero2007optical} closely. The derivation is repeated here to show that the use of a separate red laser to read out the motion does not affect the amplitude response.
We assume there is a weakly intensity-modulated laser actuating the graphene membrane, the force induced by this laser $F$ can be written as:
\begin{equation}
F(z(t),t) = (1+\varepsilon(t)) F_{\mathrm{ph}}(z(t)),
\end{equation}
where $F_{\mathrm{ph}}$ is a photo-induced force (which can be photo-thermal, radiation pressure or radiometric pressure) that is exerted on the compliant graphene membrane and $\varepsilon$ a modulation parameter assumed to be much smaller than 1. In general, the graphene membrane will show a delayed response in its deflection due to this force, but responds with a certain time delay $\tau$. This time-delay can be described by a function $h(t-t')$ that leads to the following formulation for the force:
\begin{equation}
F(z(t),t) = F_{\mathrm{ph}}(z_0) + \int_0^t \left( \frac{\partial F_{\mathrm{ph}}}{\partial t'} + \frac{\partial F_{\mathrm{ph}}}{\partial z} \frac{\partial z}{\partial t'}\right) h(t-t') \mathrm{d}t'. 
\end{equation}
In the case of constant illumination, this equation becomes:
\begin{equation}
F(z(t),t) = F_{\mathrm{ph}}(z_0) + \int_0^t \frac{\partial F_{\mathrm{ph}}}{\partial z} \frac{\partial z}{\partial t'} h(t-t') \mathrm{d}t',
\end{equation}
which is valid for the red laser used in the interferometric detection. The combined action of the red and blue laser gives for the total force:
\begin{eqnarray}
F_{\mathrm{tot}}(z(t),t) = F_{\mathrm{ph,blue}}(z_0) +  F_{\mathrm{ph,red}}(z_0) + \nonumber \\ \int_0^t \left( \frac{\partial F_{\mathrm{ph,blue}}}{\partial t'} + \frac{\partial F_{\mathrm{ph,blue}}}{\partial z} \frac{\partial z}{\partial t'}+ \frac{\partial F_{\mathrm{ph,red}}}{\partial z} \frac{\partial z}{\partial t'}\right) h(t-t') \mathrm{d}t'. 
\end{eqnarray}
The equation of motion that needs to be solved now reads:
\begin{eqnarray}\label{eq:motion1}
m \ddot{z} (t) + m \zeta \dot{z} (t) + K z(t) = \nonumber \\ 
F(z_0) +  \int_0^t \left( \frac{\partial F_{\mathrm{ph,blue}}}{\partial t'} + \frac{\partial F_{\mathrm{ph,blue}}}{\partial z} \frac{\partial z}{\partial t'} + \frac{\partial F_{\mathrm{ph,red}}}{\partial z} \frac{\partial z}{\partial t'}\right) h(t-t') \mathrm{d}t',
\end{eqnarray}
where $z(t)$ is the amplitude on a generalized coordinate on the membrane, which we assume is the amplitude detected by the red laser; $m$ is the effective mass, $\zeta$ the effective damping coefficient and $K$ the effective stiffness. We drop the term $F(z_0)$, since this leads only to a static deflection with no time-dependence. We assume that the amplitude is very small and that the terms $\partial F_{\mathrm{ph}} / \partial z$ can be approximated by a constant $\nabla F$.
  Using the properties of Laplace transforms for convolutions we can now write eq. \ref{eq:motion1} in the frequency domain:
\begin{equation}
-\omega^2 m z_{\omega} + i \omega m \zeta z_{\omega} + K z_{\omega} = i\omega F_{\mathrm{ph,blue}} h_{\omega} + \nabla F_{\mathrm{blue}} i\omega z_{\omega} h_{\omega} + \nabla F_{\mathrm{red}} i \omega z_{\omega} h_{\omega}
\end{equation}
We assume the shape of the delay function is of the exponential type:
\begin{equation}\label{eq:exponentialtype}
h(t) = 1 - \mathrm{e}^{-t/\tau},
\end{equation}
which has the Laplace transform:
\begin{equation}
h_{\omega} = \frac{1}{i \omega (1+ i \omega \tau)}.
\end{equation}
Inserting this into the equation of motion gives:
\begin{equation}
-\omega^2 m z_{\omega} + i \omega m \zeta z_{\omega} + K z_{\omega} =\frac{ F_{\mathrm{ph,blue}}}{1+i \omega \tau}  + \frac{\nabla F_{\mathrm{blue}}}{1+ i \omega \tau} z_{\omega} + \frac{ \nabla F_{\mathrm{red}}}{1 + i \omega \tau} z_{\omega}
\end{equation}.
Now we split the effective actuation force into $F_{\mathrm{ph,blue}} = A_{\mathrm{ph,blue}} \varepsilon_{\omega}$ where $A_{\mathrm{ph,blue}}$ represents the force from the DC blue laser power and  $\varepsilon_{\omega}$ is the modulation parameter. $\varepsilon_{\omega}$ is made constant as function of frequency by the calibration method described in section S3.  Regrouping in terms of omega gives:
\begin{equation}
- \omega^2 m z_{\omega} + i \omega \left[m \zeta + \frac{ \tau \nabla F}{1+\omega^2 \tau^2} \right] z_{\omega} + \left[K - \frac{ \nabla F}{1+\omega^2 \tau^2} \right] z_{\omega} = \frac{A_{\mathrm{ph,blue}} \varepsilon_{\omega}}{1+ i\omega \tau}, 
\end{equation}
where we combined $\nabla F_{\mathrm{blue}}$ and $\nabla F_{\mathrm{red}}$ into $\nabla F = \nabla F_{\mathrm{blue}} + \nabla F_{\mathrm{red}}$. The solution for the amplitude is:
\begin{equation}
z_{\omega} = \frac{A_{\mathrm{ph,blue}} \varepsilon_{\omega}} { (-m\omega^2 + i \omega m \zeta +k) (1+i\omega \tau) - \nabla F},
\end{equation} 
with real and imaginary part:
\begin{eqnarray}
\mathcal{R}(z_{\omega}) = A_{\mathrm{ph,blue}} \varepsilon_{\omega} \frac{-m \omega^2 - m \zeta \omega^2 \tau + K - \nabla F}{(-m\omega^2 - m \zeta \omega^2 \tau + k \nabla F)^2 + (\tau \omega (k-m\omega^2) + m \zeta \omega)^2},\\
\mathcal{I}(z_{\omega}) = A_{\mathrm{ph,blue}} \varepsilon_{\omega} \frac{- \tau \omega (k-m\omega^2) - m \zeta \omega}{(-m\omega^2 - m \zeta \omega^2 \tau + k \nabla F)^2 + (\tau \omega (k-m\omega^2) + m \zeta \omega)^2}.
\end{eqnarray}
In the limit where $1/\tau \ll \sqrt{k/m}$ the imaginary part shows a local maximum at a radial frequency of approximately $1/\tau$. In this case, the frequency response function is $\omega \tau/(1 + \omega^2 \tau^2)$, which is used in this work to fit to the imaginary part of the measured response.

\section*{Derivations of interfacial thermal resistance, specific heat and characteristic thermal time for 2D materials}

\subsection*{S7: Derivation of expression for $\tau$ (eq. 8 in main text)}
The interfacial thermal resistance $R_B$ can be determined by using the heat flux:
\begin{equation}\label{eq:resdef}
R_B = \frac{A \Delta T}{Q},
\end{equation}
where $A$ is cross sectional area of the boundary, $\Delta T$ the temperature difference and $Q$ the heat flux. The first step in determining the interfacial resistance, is thus to determine the heat flux that crosses the interface. The heat flux that crosses from interface 1 to interface 2 can be expressed by \cite{gray1981nonequilibrium,peterson1973kapitza}:
\begin{equation}\label{eq:simpleheatflux}
Q_{1\rightarrow2} = U \nu \bar{w},
\end{equation}
where $U$ is the total energy per unit volume of the heat carriers, $\nu$ the velocity at which they propagate and $\bar{w}$ the probability that the heat carriers transmit over the interface. In the calculation of thermal interfacial resistance, the difficulty lies in calculating the transmission probability $\bar{w}$, while the calculation of energy and propagation velocity is quite straigtforward. 

Our approach is thus, to calculate the energy and velocity and use that to estimate the value of $\bar{w}$ from the measurement. In order to do this, it is assumed that the heat in graphene is carried by phonons and that all the heat is carrier by three acoustic phonon polarizations, the longitudinal (LA), transverse (TA) and flexural (ZA). The LA and TA branch are far below the Debye temperature of 2100 K due to their large velocities, but the ZA branch will be fully thermalized since its Debye temperature is at 50 K \cite{hone2001phonons}. The contribution to the heat flux of each polarization can be added to obtain:
\begin{equation}
Q_{1\rightarrow2} = \sum\limits_j U_{j} \nu_{j} \bar{w}_{j},
\end{equation}
and the total heat flux becomes:
\begin{equation}\label{eg:heatfluxtot1}
Q_{1\rightarrow2} -Q_{2\rightarrow1}  = \sum\limits_j U_{1j} \nu_{1j} \bar{w}_{1j}- \sum\limits_j U_{2j} \nu_{2j} \bar{w}_{2j},  
\end{equation}
the index $ij$ now describes the material ($i = 1$ for suspended, $i =2$ for supported graphene) and phonon mode $j$. 

To calculate the energy $U_{ij}$, we start from the Bose-Einstein distribution to find the average phonon number $ \langle n (\omega) \rangle$ at a fixed frequency $\omega$:
\begin{equation}
\langle n (\omega) \rangle = \frac{1}{\mathrm{e}^{\hbar \omega/k_B T} -1},
\end{equation}
where $\hbar$ is the reduced Plack's constant, $k_B$ the Boltzmann constant and $T$ temperature. The energy carried by each phonon is $\hbar \omega$, therefore we can write the average energy $\langle E(\omega) \rangle$ that phonons have at this frequency:
\begin{equation}
\langle E(\omega) \rangle = \hbar \omega \langle n \rangle =  \frac{\hbar \omega}{\mathrm{e}^{\hbar \omega/k_B T} -1}.
\end{equation}
The number of states (density of states $D(\omega)$) that is accessible to the system between the frequencies $\omega$ and $\omega+ \mathrm{d} \omega$ is defined by:
\begin{equation}
\mathrm{d}N_{ij} = D_{ij}(\omega) \mathrm{d} \omega
\end{equation}
and this makes the total energy $V_{\ij}$ in the system:
\begin{equation}
V_{ij} = \int_0^{\infty} \mathrm{d} \omega D_{ij}(\omega) \langle E(\omega) \rangle= \int_0^{\infty} \mathrm{d} \omega D_{ij}(\omega) \frac{\hbar \omega}{\mathrm{e}^{\hbar \omega/k_B T} -1}.
\end{equation}
Next, one has to know for the density of states, how many modes are available in the momentum space. In a 2-dimensional crystal, if we know the size of the system $A$, the uncertainty in momentum is $(2 \pi \hbar)^2/A$ and the number of modes available to the system becomes:
\begin{equation}
N = A \int \frac{\mathrm{d}^2 p}{(2\pi \hbar)^2} = A \int \frac{\mathrm{d}^2 k}{(2 \pi)^2},
\end{equation}
where one can integrate over circles with circumference $2 \pi k$ to obtain:
\begin{equation}
N = A \int \frac{ k \mathrm{d} k}{2 \pi}.
\end{equation}
Using $\mathrm{d} N = D(\omega) \mathrm{d} \omega$ we obtain for the total energy in the system:
\begin{equation}
\mathrm{d} V_{ij} (\omega) \mathrm{d}\omega = A \frac{k \mathrm{d} k}{2\pi} \frac{\hbar \omega}{\mathrm{e}^{\hbar \omega/k_B  T} -1},
\end{equation}
which is divided by the total volume of the system to obtain for $U_{ij}$:
\begin{equation}\label{eq:endens}
\mathrm{d} U_{ij} (\omega) \mathrm{d}\omega =  \frac{k \mathrm{d} k}{2\pi h_g} \frac{\hbar \omega}{\mathrm{e}^{\hbar \omega/k_B  T} -1},
\end{equation}
to perform the integration, it is necessary to use the dispersion relation that relates the frequency to the wavenumber. Since the flexural phonons have different properties than the transverse and longitudinal phonon, these will have to be analyzed separately in the sections below. 

\subsubsection{Longitudinal and tranverse modes}
For the longitudinal and transverse acoustic phonons, we can write the linear dispersion relationship:
\begin{equation}
\omega = c_{ij} k,
\end{equation}
where $c_{ij}$ is the propagation velocity of the phonons, since $\nu = \frac{\mathrm{d}\omega}{\mathrm{d}k} = c_{ij}$. Substitution into eq. \ref{eq:endens} gives:
\begin{equation}\label{eq:endenslin}
\mathrm{d} U_{ij} (\omega) \mathrm{d}\omega =  \frac{ \hbar  \omega^2 \mathrm{d} \omega}{2\pi c_{ij}^2 h_g} \frac{1}{\mathrm{e}^{\hbar \omega/k_B  T} -1},
\end{equation}
Using this, we can write for the heat flux over an interface of area $A$ from material 1 to material 2:
\begin{equation}
Q_{1\rightarrow2, j} = {A  U_{1j} \nu_{1j} \bar{w}_{1j}} =\int_0^{\omega_D}   \frac{ A \bar{w}_{1j} \hbar  \omega^2 \mathrm{d} \omega}{2\pi c_{1j}  h_g} \frac{1}{\mathrm{e}^{\hbar \omega/k_B  T} -1}.
\end{equation}
To solve the frequency integral, we assume $k_B T \ll  \hbar \omega_D$ and using a coordinate transform $x = \hbar \omega/k_B T$:
\begin{equation}
Q_{1\rightarrow2, j} =\int_0^{\infty}   \frac{A \bar{w}_{1j} k_B^3 T_1^3 }{2\pi  \hbar^2 c_{1j}  h_g} \frac{x^2 \mathrm{d} x}{\mathrm{e}^{x} -1}
\end{equation}
We assume that the transmission probability frequency-independent. This results in:
\begin{equation}
Q_{1\rightarrow2, j} = \frac{ A \zeta(3) \bar{w}_{1j} k_B^3 T_1^3}{\pi \hbar^2 c_{1j}  h_g} 
\end{equation}
and for the heat flux from material 2 to material 1:
\begin{equation}
Q_{2\rightarrow1, j} = A U_{2j} \nu_{2j} \bar{w}_{2j} =\frac{ A \zeta(3) \bar{w}_{2j} k_B^3 T_2^3}{\pi  \hbar^2 c_{2j}  h_g}. 
\end{equation}

\subsubsection{Flexural mode}
When strain is present in a 2D lattice, the dispersion of the flexural phonons can be written as \cite{lifshitz1952thermal}:
\begin{equation}
\omega^2 = s_{iz}^2 k^4 + c_{iz}^2 k^2
\end{equation}
which has 4 solutions for $k$, however implying the conditions $s_{iz} > 0$, $c_{iz} > 0$, $k > 0$ and enforcing that $k$ must be a real number, we only have one solution:
\begin{equation}\label{eq:parabolick}
k = \frac{1}{\sqrt{2}} \sqrt{\frac{\sqrt{c_{iz}^4 + 4 s_{iz}^2 \omega^2}}{s_{iz}^2} - \frac{c_{iz}^2}{s_{iz}^2}},
\end{equation}
this can be substituted in eq. \ref{eq:endens} to obtain:
\begin{equation}
\mathrm{d} U_{iz} (\omega) \mathrm{d}\omega =  \frac{\hbar \omega^2 \mathrm{d} \omega}{2\pi h_g\sqrt{c_{iz}^4 + 4 s_{iz}^2 \omega^2}} \frac{1}{\mathrm{e}^{\hbar \omega/k_B  T} -1}
\end{equation}
Note, that this equation converges either to the expression for linear dispersion if $c_{iz}^4 \gg 4s_{iz}\omega^2$ or to the expression for quadratic dispersion if $c_{iz}^4 \ll 4s_{iz}^2\omega^2$. The propagation velocity becomes:
\begin{equation}
\nu = \frac{\mathrm{d}\omega}{\mathrm{d}k} = \frac{ \mathrm{d}\sqrt{s_{iz}^2 k^4 + c_{iz}^2 k^2}}{\mathrm{d} k} = \frac{c_{iz}^2 k + 2 s_{iz}^2 k^3}{\sqrt{c_{iz}^2 k^2 + s_{iz}^2 k^4}} = \frac{c_{iz}^2 + 2 s_{iz}^2 k^2}{\sqrt{c_{iz}^2 + s_{iz}^2 k^2}}
\end{equation} 
substituting eq. \ref{eq:parabolick} gives:
\begin{equation}
\nu_{iz} = \frac{\sqrt{2} \sqrt{c_{iz}^4 + 4 s_{iz}^2 \omega^2}}{\sqrt{c_{iz}^2 + \sqrt{c_{iz}^4 + 4 s_{iz}^2 \omega^2}}} ,
\end{equation}
which in the limit case of purely quadratic dispersion $c_{iz}=0$ becomes $\nu_{iz} = 2\sqrt{\omega s_{iz}}$ and in the case of linear dispersion $s_{iz} = 0$ becomes $\nu_{iz} = c_{iz}$. 

The analysis can be simplified by assuming either high or low strains, which should follow from our experiments. It can be seen, that the condition:
 \begin{equation}
 \frac{2 k_B T s_{iz}}{\hbar} \gg c^2
 \end{equation}
 allows us to use descibe the heat transport of the ZA branch by a quadratic dispersion without strain, while the condition:
 \begin{equation}\label{eq:llcondition}
  \frac{2 k_B T s_{iz}}{\hbar} \ll c^2
\end{equation}
allows one to use a linear dispersion for the ZA branch. From the resonance frequencies in our experiments, we can estimate the strain present in the drum resonators:
\begin{equation}
\epsilon = \frac{\rho h \omega^2 a^2}{E h 2.4048^2 },
\end{equation}
assuming $\rho h = \num{7.7e-7}$ kg/m$^2$ and $E h =$ 340 N/m, we find the lowest observed strain $\epsilon_{\mathrm{low}} = \num{1.026e-5}$ and the highest observed strain $\epsilon_{\mathrm{high}} = \num{1.71e-4}$. Expressions for coefficients $s_{iz}$ and $c_{iz}$ are given by Lifshitz \cite{lifshitz1952thermal}:
\begin{equation}
s_{iz} = \sqrt{\frac{\kappa}{\rho}} 
\end{equation}
where $\kappa$ is the bending rigidity, which is $\kappa = \num{1e-19}$ J for single-layer graphene, and:
\begin{equation}
c_{iz}= \sqrt{2u\frac{\lambda+\mu}{\rho}}
 \end{equation}
 where $u$ is the dilatation and $\lambda$, $\mu$ are the Lame parameters. Now we can calculate the coefficients for the lowest strain:
 \begin{equation}
  \frac{4 k_B^2 T^2 s_{iz}^2}{\hbar^2} = \num{2.6828e+05},
  \end{equation} 
\begin{equation}
c_{iz}^4 = \num{1.3683e+08}
\end{equation}
from which we conclude that for each drum measured in this work the condition in eq. \ref{eq:llcondition} holds. Due to the low velocities the Debye temperature of the flexural phonons is much lower than the in-plane phonons. Therefore, we have to write the heat flux as:
 \begin{equation}
Q_{1 \rightarrow 2,z} = \int_0^{\omega_D} \frac{A \bar{w}_{1z} \hbar \omega^2 \mathrm{d} \omega}{2 \pi c_{1z} h_g} \frac{1}{\mathrm{e}^{\hbar \omega/k_B T} -1} = \int_0^{\theta/T} \frac{A \bar{w}_{1z} k_B^3 T^3 \mathrm{d} x}{2 \pi c_{1z} \hbar^2 h_g} \frac{x^2}{\mathrm{e}^x -1},  
 \end{equation}
 where $\theta$ is the Debye temperature, since we are above the Debye temperature $e^x \approx 1+x$:
\begin{equation}
Q_{1 \rightarrow 2,z} =  \int_0^{\theta/T} \frac{A \bar{w}_{1z} k_B^3 T^3 x \mathrm{d} x}{2 \pi c_{1z} \hbar^2 h_g} 
=\frac{A \bar{w}_{1z} k_B^3 T}{4 \pi c_{1z} \hbar^2 h_g} {\theta^2},
\end{equation}
the total number of states in the system is:
\begin{equation}
N = A_g \int_0^{\omega_D} \frac{\omega \mathrm{d}\omega}{2 \pi c_{1z}^2} = A_g \frac{\omega_D^2}{4 \pi c_{1z}^2}
\end{equation}
the Debye temperature becomes:
\begin{equation}
\theta = \frac{\hbar \omega_D}{k_B} =  \frac{\hbar}{k_B} \sqrt{\frac{4 \pi c^2 N}{\pi a^2}},
\end{equation}
here $N/\pi a^2$ is the number of states per unit square, which is limited by the area of the unit cell $A_{uc} = \num{5e-20}$ m$^2$. The heat flux from the ZA mode now becomes:
\begin{equation}
Q_{1\rightarrow2} = \frac{A \bar{w}_{1z} k_B T c_{1z}}{h_g A_{uc}}
\end{equation}
and the total heat flux now becomes:
\begin{eqnarray}
Q_{tot} =\frac{ \zeta(3) A k_B^3 T_1^3 }{\pi  \hbar^2   h_g} \left( \frac{\bar{w}_{1l}}{c_{1l}} + \frac{\bar{w}_{1t}}{c_{1t}}  \right) + \frac{A \bar{w}_{1z} k_B T_1 c_{1z}}{h_g A_{uc}}\nonumber \\    
  -  \frac{\zeta(3) A k_B^3 T_2^3 }{\pi  \hbar^2 h_g} \left( \frac{\bar{w}_{2l}}{c_{2l}} + \frac{\bar{w}_{2t}}{c_{2t}}  \right)  -\frac{A \bar{w}_{2z} k_B T_2 c_{2z}}{h_g A_{uc}}\nonumber \\
  =\frac{ \zeta(3) A k_B^3 T_1^3 }{\pi  \hbar^2   h_g} \left( \frac{\bar{w}_{1l}}{c_{1l}} + \frac{\bar{w}_{1t}}{c_{1t}} +\frac{\pi \hbar^2 \bar{w}_{1z} c_{1z}}{k_B^2 T_1^2 \zeta(3) A_{uc}}  \right) \nonumber \\    
  -  \frac{\zeta(3) A k_B^3 T_2^3 }{\pi  \hbar^2 h_g} \left( \frac{\bar{w}_{2l}}{c_{2l}} + \frac{\bar{w}_{2t}}{c_{2t}} +\frac{\pi \hbar^2 \bar{w}_{2z} c_{2z}}{k_B^2 T_2^2 \zeta(3) A_{uc}} \right)  
\end{eqnarray}
The condition $Q_{tot} = 0$ has to apply if $T_1 = T_2$, this implies that:
\begin{equation}
\frac{\bar{w}_{1l}}{c_{1l}} + \frac{\bar{w}_{1t}}{c_{1t}} +\frac{\pi \hbar^2 \bar{w}_{1z} c_{1z}}{k_B^2 T^2 \zeta(3) A_{uc}}  = \frac{\bar{w}_{2l}}{c_{2l}} + \frac{\bar{w}_{2t}}{c_{2t}} +\frac{\pi \hbar^2 \bar{w}_{2z} c_{2z}}{k_B^2 T^2 \zeta(3) A_{uc}}
 \end{equation} 
 which makes the heat flux:
 \begin{equation}
 Q_{tot} =\frac{ \zeta(3) A k_B^3 (T_1^3  - T_2^3)}{\pi  \hbar^2   h_g} \left( \frac{\bar{w}_{1l}}{c_{1l}} + \frac{\bar{w}_{1t}}{c_{1t}} +\frac{\pi \hbar^2 \bar{w}_{1z} c_{1z}}{k_B^2 T_1^2 \zeta(3) A_{uc}}  \right) 
 \end{equation}
 This can be linearized for small temperature differences $\Delta T$ to obtain:
 \begin{equation}
 Q_{tot} =\frac{ 3 \zeta(3) A k_B^3 T^2 \Delta T }{\pi  \hbar^2 h_g}  \left( \frac{\bar{w}_{1l}}{c_{1l}} + \frac{\bar{w}_{1t}}{c_{1t}} +\frac{\pi \hbar^2 \bar{w}_{1z} c_{1z}}{k_B^2 T^2 \zeta(3) A_{uc}}  \right) 
 \end{equation}
 and the boundary resistance is directly obtained from eq. \ref{eq:resdef}:
 \begin{equation}\label{eq:resistancetotal}
R_B=\frac{ \pi  \hbar^2 h_g  }{  3 \zeta(3)  k_B^3 T^2}   \left( \frac{\bar{w}_{1l}}{c_{1l}} + \frac{\bar{w}_{1t}}{c_{1t}} +\frac{\pi \hbar^2 \bar{w}_{1z} c_{1z}}{k_B^2 T^2 \zeta(3) A_{uc}}  \right)^{-1}
 \end{equation}

\subsubsection{Specific heat}
We can also calculate the specific heat $c_p$ by starting from equation \ref{eq:endenslin}:
 \begin{equation}
\mathrm{d} U_{ij} (\omega) \mathrm{d}\omega =  \frac{ \hbar  \omega^2 \mathrm{d} \omega}{2\pi c_{ij}^2 h_g} \frac{1}{\mathrm{e}^{\hbar \omega/k_B  T} -1},
\end{equation}
\begin{equation}
U_{1j} =  \int_0^{\omega_D}\frac{ \hbar  \omega^2 \mathrm{d} \omega}{2\pi c_{ij}^2 h_g} \frac{1}{\mathrm{e}^{\hbar \omega/k_B  T} -1}
=  \int_0^{\infty} \frac{x^2 k_B^3 T^3 \mathrm{d} x}{2\pi c_{ij}^2 \hbar^2 h_g} \frac{1}{\mathrm{e}^{x} -1}
 =  \frac{\zeta(3) k_B^3 T^3}{\pi c_{ij}^2 \hbar^2 h_g},
 \end{equation}
 which is valid for the LA and TA branches. For the ZA phonons we have to take the high temperature limit:
 \begin{equation}
 U_{1z} =  \int_0^{\theta/T} \frac{x^2 k_B^3 T^3 \mathrm{d} x}{2\pi c_{ij}^2 \hbar^2 h_g} \frac{1}{\mathrm{e}^{x} -1},
 \end{equation}
 using $\mathrm{e}^x \approx 1+x$:
\begin{equation}
 U_{1z} =  \int_0^{\theta/T} \frac{x k_B^3 T^3 \mathrm{d} x}{2\pi c_{ij}^2 \hbar^2 h_g} = \frac{k_B^3 T}{4 \pi c_{1z}^2 \hbar^2 h_g} {\theta^2},
 \end{equation}
 now by substituting:
 \begin{equation}
   \theta_{1z} = \frac{\hbar}{k_B} \sqrt{\frac{4\pi c_{1z}^2}{A_{uc}}},
\end{equation}
 the energy density becomes:
 \begin{equation}
 U_{1z} = \frac{k_B T}{h_g A_{uc}}
 \end{equation}

  \begin{equation}
U_1= \sum\limits_j U_{1j} =  \frac{\zeta(3) k_B^3 T^3}{\pi  \hbar^2 h_g} \left( \frac{1}{c_{1l}^2} +  \frac{1}{c_{1t}^2}\right) + \frac{k_B T}{h_g A_{uc}} ,
 \end{equation}
 Now we find:
 \begin{equation}
 \rho c_p = \frac{\mathrm{d} U}{\mathrm{d} T} = \frac{3 \zeta(3) k_B^3 T^2}{\pi  \hbar^2 h_g}  \left( \frac{1}{c_{1l}^2} +  \frac{1}{c_{1t}^2}\right) +\frac{k_B}{h_g A_{uc}},   
 \end{equation}
 \begin{equation}\label{eq:cp}
 c_p   =  \frac{3 \zeta(3) k_B^3 T^2}{\pi   \rho \hbar^2 h_g}  \left( \frac{1}{c_{1l}^2} +  \frac{1}{c_{1t}^2} + \frac{\pi \hbar^2}{3 \zeta(3) A_{uc} k_B^2 T^2}\right)
 \end{equation}
In the main text we found the expression:
 \begin{equation}
 \tau = \frac{\rho c_p a}{2} R_B
 \end{equation}
 Substituting eqs. \ref{eq:resistancetotal} and \ref{eq:cp} gives:
 \begin{equation}
 \tau = \frac{a}{2} \frac{ \dfrac{1}{c_{1l}^2} +  \dfrac{1}{c_{1t}^2} + \dfrac{\pi \hbar^2}{3 \zeta(3) A_{uc} k_B^2 T^2}}{\dfrac{\bar{w}_{1l}}{c_{1l}} + \dfrac{\bar{w}_{1t}}{c_{1t}} +\dfrac{\pi \hbar^2 \bar{w}_{1z} c_{1z}}{k_B^2 T^2 \zeta(3) A_{uc}}},
 \end{equation}
this result was used in the main text to estimate the average phonon transmission probability $\bar{w}$. This average is defined as the situation where all branches have equal transmission probability: $\bar{w} = \bar{w}_{1t} = \bar{w}_{1l} = \bar{w}_{1z}$.

\end{document}